\documentclass[conference]{IEEEtran}

\IEEEoverridecommandlockouts
\usepackage{epsfig}
\usepackage{graphicx}
\usepackage{amsmath,amssymb,amsfonts}
\usepackage{algorithmic}
\usepackage{graphicx}
\usepackage{textcomp}
\usepackage{xcolor}
\usepackage{multirow}
\usepackage{url} 
\usepackage{booktabs}
\usepackage{fancyhdr}
\usepackage{flushend}
\def\BibTeX{{\rm B\kern-.05em{\sc i\kern-.025em b}\kern-.08em
    T\kern-.1667em\lower.7ex\hbox{E}\kern-.125emX}}

\usepackage{mathtools,subcaption}

\fancypagestyle{firstpage}{
  \setlength{\headheight}{3.5\normalbaselineskip}
  \setlength{\headsep}{1.5\normalbaselineskip}

  \fancyhead[C]{%
    \footnotesize%
    \textit{Proc. IEEE MMSP 2024}, Purdue University, IN, USA, Oct. 2024 \\
    \vspace{0.8\normalbaselineskip}%
    \scriptsize%
    \copyright{} 2024 IEEE. Personal use of this material is permitted. Permission from IEEE must be obtained for all other uses, in any current or future media, including reprinting/republishing this material for advertising or promotional purposes, creating new collective works, for resale or redistribution to servers or lists, or reuse of any copyrighted component of this work in other works.%
  }
  \fancyfoot{}
}

\begin{document}

\title{Learned Multimodal Compression for Autonomous Driving}

\author{Hadi Hadizadeh and Ivan V. Baji\'{c} \thanks{This work was supported in part by Digital Research Alliance of Canada, and NSERC grants RGPIN-2021-02485 and RGPAS-2021-00038.}\\
Simon Fraser University, Burnaby, BC, Canada \\
{\tt\small hadi\_hadizadeh@sfu.ca, ibajic@ensc.sfu.ca}
}

\maketitle

\thispagestyle{firstpage}

\begin{abstract}
Autonomous driving sensors generate an enormous amount of data. In this paper, we explore learned multimodal compression for autonomous driving, specifically targeted at 3D object detection. We focus on camera and LiDAR modalities and explore several coding approaches. One approach involves joint coding of fused modalities, while others involve coding one modality first, followed by conditional coding of the other modality. We evaluate the performance of these coding schemes on the nuScenes dataset. Our experimental results indicate that joint coding of fused modalities yields better results compared to the alternatives.   
\end{abstract}

\vspace{5pt}


\section{Introduction}
\label{sec:intro}
With the rapid advancement of artificial intelligence technologies, multimodal learning has emerged as a powerful paradigm for data processing and analysis in various applications, including autonomous robots and vehicles~\cite{bojarski2016end}, and large language models~\cite{bubeck2023sparks}. By integrating data from various modalities, multimodal learning enhances the robustness and adaptability of AI systems, resulting in improved performance and a greater capacity to effectively address real-world challenges.

A notable application of multimodal learning is in autonomous driving, where it enables autonomous vehicles (AVs) to interpret and respond to their environment using multiple sources of data, such as cameras, LIDAR, radar, and ultrasonic sensors. This integration of diverse data types allows for more accurate machine perception and decision making, enhancing the vehicle's ability to detect and understand objects, predict their movements, and navigate complex driving scenarios safely. Using the information contained in different modalities, AVs can achieve greater robustness and reliability, reducing the risk of accidents, and improving overall performance in varied and dynamic driving conditions.

AV sensors generate a huge amount of data~\cite{Gotz_data_AV}, which, depending on sensor configuration, could reach 40 Gigabits per second (Gbps). These data must be processed and analyzed in real time, which requires a large amount of energy and, in turn, reduces the range of the vehicle~\cite{Richart_energy_AV}. Reducing the amount of data is therefore important, whether for onboard processing or for computational offloading to cloud services using future low-latency communications~\cite{Gupta_low_latency_6G}. Hence, efficient compression techniques are essential to reduce the amount of sensory data while preserving the critical information necessary for accurate functioning and decision-making. This is a case for \emph{coding for machines}, as the data from the various sensors is intended for autonomous decision-making, without human involvement.  

Although data compression research has been experiencing a renaissance in recent years due to the emergence of learning-based coding strategies~\cite{balle2018, minnen, cheng2020, canf, anfic}, there has been limited work on sensor data compression for AVs. Within this domain, the effect of camera data compression by Motion JPEG (MJPEG) on 2D object detection in AVs has been studied in ~\cite{Camera_coding_AV_LIM_2021}. Further, LiDAR compression strategies for AVs have been surveyed in~\cite{LiDAR_coding_AV}, but their impact on the resulting scene analysis has not been studied.  Meanwhile,~\cite{Data_compression_AV_2021} looked at both camera and LiDAR data compression using the same coding pipeline, but only considered separate (uni-modal) compression strategies. This, however, ignores the considerable amount of inter-modal redundancy that exist in AV sensory data~\cite{Multimodal_MIPR_2024}.

In this paper, we present a learned multimodal compression system for 3D object detection in autonomous driving.  We focus on camera and LiDAR, as the two most data-intensive modalities~\cite{Gotz_data_AV}, and develop several approaches following the coding-for-machines principles~\cite{hyomin_sic,Alon_RDCFM}. In other words, we consider not only removing rendundancy (intra- and inter-modal), but also task-irrelevant information. One coding approach we consider involves joint coding of fused modalities, while others involve coding
one modality first, followed by conditional coding of the other modality. 
To the best of our knowledge, this is the first exploration of multimodal coding for machines for autonomous driving in the literature.

The paper is organized as follows. Preliminaries and motivation for the proposed coding approaches are provided in Section~\ref{sec:motivation}, followed by the descriptions of the methods themselves in Section~\ref{sec:proposed}. The experimental results are provided in Section~\ref{sec:experiments}, and the conclusions are drawn in Section~\ref{sec:conclusions}.

\section{Preliminaries and Motivation}
\label{sec:motivation}
Let $X$ be the input signal. Conventional codecs usually follow the autoencoder processing chain $X \to \mathcal{Y} \to \widehat{X}$, where $\mathcal{Y}$ is the transformed and encoded representation of the input, and $\widehat{X}\approx X$. The design of conventional codecs is based on rate-distortion optimization
\begin{equation}
    \min_{p(y | x)} \quad H(\mathcal{Y}) + \lambda \cdot \mathbb{E}[d(X; \widehat{X})],
    \label{eq:RD}
\end{equation}
where $p(y|x)$ is the encoding function, $H(\cdot)$ is the entropy~\cite{Cover_Thomas_2006}, $d(\cdot, \cdot)$ is a distortion measure, like Mean Squared Error (MSE), Structural Similarity Index Metric (SSIM), etc., $\mathbb{E}[\cdot]$ is the expectation operator, and $\lambda$ is the Lagrange multiplier. 

On the other hand, codecs for machines usually follow the processing chain $X \to \mathcal{Y} \to \widehat{T}$, where $\widehat{T}$ is a decision, such as a class label, a set of object detection boxes, etc. In other words, the goal of coding for machines is not to recover the input signal, but to provide sufficient information for making the necessary decision. The design of codecs for machines usually follows the Information Bottleneck~\cite{IB_Allerton1999} principle:
\begin{equation}
    \min_{p(y | x)} \quad I(X; \mathcal{Y}) - \beta \cdot I(\mathcal{Y}; \widehat{T}),
\label{eq:IB}
\end{equation}
where $I(\cdot;\cdot)$ is the mutual information~\cite{Cover_Thomas_2006} and and $\beta$ is the Lagrange multiplier.. Comparing equations~(\ref{eq:RD}) and~(\ref{eq:IB}) reveals the key differences between conventional codecs and codecs for machines. Conventional codecs try to remove statistical redundancy (minimize $H(\mathcal{Y})$) while keeping the recovered signal $\widehat{X}$ close to the orignal input $X$. Meanwhile, codecs for machines try to remove all information from the input (minimize $I(X; \mathcal{Y})$) except that which is related to the task (maximize $I(Y; \widehat{T})$). In other words, besides removing statistical redundancy, which is implicit in the minimization of $I(X; \mathcal{Y})$, codecs for machines \emph{also try to remove task-irrelevant information}. 

\begin{figure*}
\centering
\includegraphics[width=0.8\textwidth]{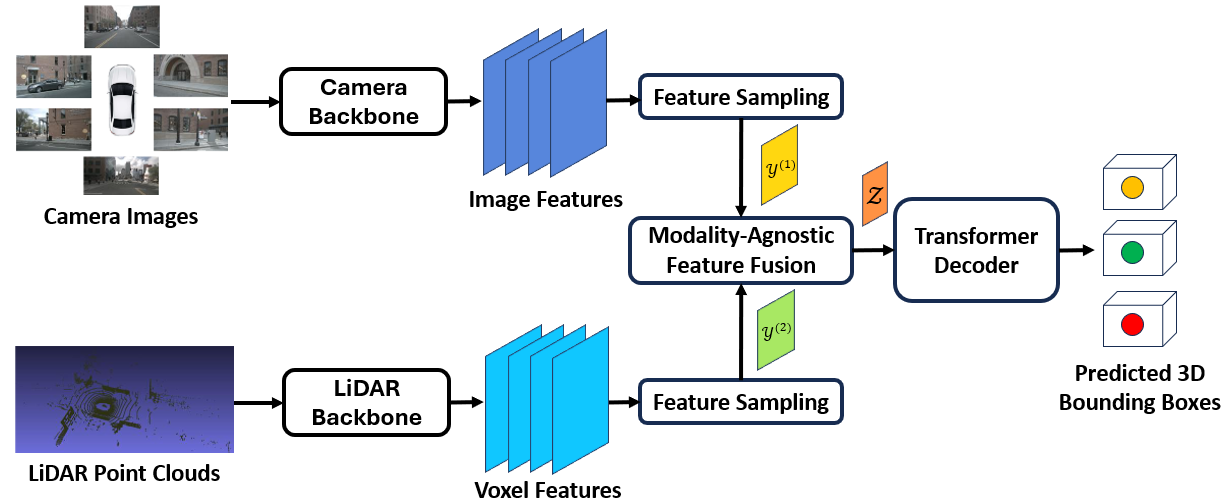}
\caption{The simplified architecture of FUTR3D. In our experiments, $\mathcal{Y}^{(1)}$ and $\mathcal{Y}^{(2)}$ are camera and LiDAR features after feature sampling, which are two single-channel feature maps of the same spatial size. $\mathcal{Z}$ is the fused modality. When using 600 queries with an embedding length of 256, these feature maps are of size $1\times 600\times 256$.}
\label{fig:futr3d}
\end{figure*}

Given a task, for example object detection, how do we know which information is irrelevant to that task? One option is to consider a feed-forward neural network trained for that task. Let $\mathcal{Y}_i$ be the features produced by the network at layer $i$, so that the processing chain implemented by the neural network is $X \to \mathcal{Y}_1 \to \mathcal{Y}_2 \to ... \to \widehat{T}$. The data processing inequality~\cite{Cover_Thomas_2006} implies that $I(X;\mathcal{Y}_i) \leq I(X;\mathcal{Y}_j)$ for $i>j$, with equality only if the mapping from $\mathcal{Y}_j$ to $\mathcal{Y}_i$ is invertible. Since typical processing implemented by network layers is not invertible,\footnote{Indeed, creating invertible neural networks is not trivial~\cite{i-ResNet_2019}.} the inequality will in general be strict. Therefore, moving down the network layers effectively implements the minimization of the first term in~(\ref{eq:IB}); in other words, it removes information about the input $X$. On the other hand, if the network is well-trained,\footnote{By ``well-trained'' we mean accurate enough for a particular application.} then the task-relevant information (the second term in~(\ref{eq:IB})) must be flowing all the way to the output. Therefore, the information that is removed by the network must be irrelevant to the task at hand. 

This provides a way to create a codec for machines: process the input by a certain number of layers from a well-trained network to remove some of the task-irrelevant information, and then encode it to remove the remaining statistical redundancy. The output of the codec is fed to the remaining layers of the network. In theory, one could encode the network's output, which does not contain any task-irrelevant information, and is therefore most efficient. However, this means having to run an entire network prior to encoding, which may be prohibitively expensive, especially for battery-operated platforms with limited computational capabilities. Therefore, in practice, running a reasonable subset of layers prior to encoding is a more feasible option.

\section{The Proposed Coding Approaches}
\label{sec:proposed}
We follow the logic presented in the previous section and employ a state-of-the-art model called FUTR3D (``Fusion Transformer for 3D Detection'')~\cite{futr3d} for multimodal object detection as a way to remove task-irrelevant information. FUTR3D is an end-to-end transformer-based sensor fusion framework for 3D object detection in autonomous driving. It provides state-of-the-art detection accuracy on nuScenes~\cite{nuscenes}, a large dataset for autonomous driving. nuScenes consists of about 34,000 data samples. Each sample includes data from three different modalities: 6 cameras (each image is of size $1600\times 900$), 5 radars, and one LiDAR sensor, accompanied by 3D bounding-box annotations.

In FUTR3D, multiple modalities (e.g. camera, LiDAR, and Radar) are fused through a query-based Modality-Agnostic Feature Sampler (MAFS), which consists of feature sampling from each modality followed by modality-agnostic feature fusion~\cite{futr3d}. The integrated data is then utilized by a transformer decoder to perform the 3D object detection task. In this work, we focus on the camera and LiDAR, as the two most data-intensive modalities~\cite{Gotz_data_AV}. Following~\cite{futr3d}, we employed ResNet-101~\cite{resnet} and VoxelNet~\cite{voxelnet} as the backbones for the camera and LiDAR modalities, respectively. Fig.~\ref{fig:futr3d} shows the simplified architecture of FUTR3D, which illustrates how the two modalities are fused together. In this figure, $\mathcal{Y}^{(1)}$ and $\mathcal{Y}^{(2)}$ are the camera and LiDAR features, respectively,  produced after feature sampling. At this point in the network, they are single-channel feature maps of the same dimension: $600 \times 256$. Much of the task-irrelevant information has been removed from them, so what is left is to encode them efficiently to remove the remaining statistical redundancy.

\begin{figure*}
\centering
\includegraphics[width=0.9\textwidth]{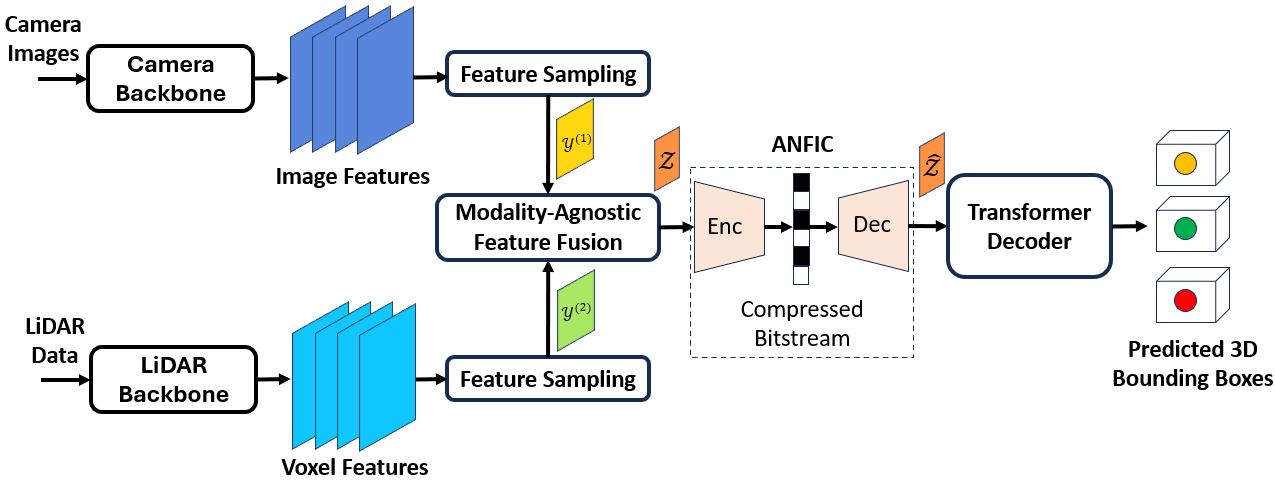}
\caption{The architecture of the first coding approach (Approach 1). In this approach, the fused modality, $\mathcal{Z}$, is first encoded by the ANFIC's encoder (Enc) to obtain a compressed bitstream. The bitstream is then decoded by the ANFIC's decoder (Dec) to reconstruct $\hat{\mathcal{Z}}$, which is then fed to the transformer decoder to perform object detection. }
\label{fig:approach1}
\end{figure*}

\begin{figure*}
\centering
\includegraphics[width=0.9\textwidth]{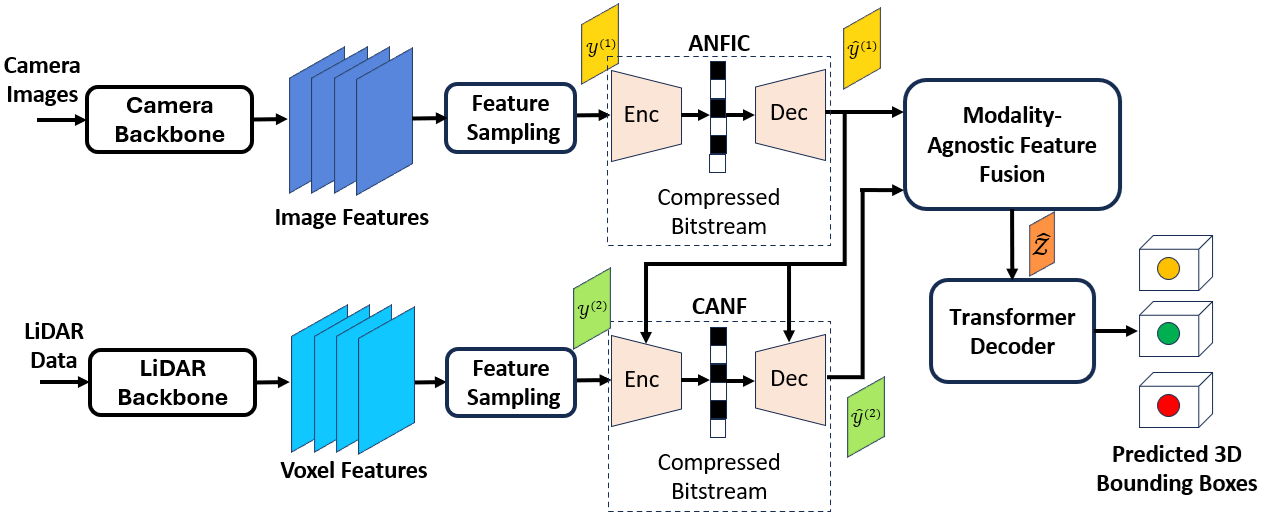}
\caption{The architecture of the second coding approach (Approach 2). In this approach, the camera modality, $\mathcal{Y}^{(1)}$, is first encoded by the ANFIC's encoder (Enc) to obtain a compressed bitstream. The bitstream is then decoded by the ANFIC's decoder (Dec) to reconstruct $\hat{\mathcal{Y}}^{(1)}$. The LiDAR modality, $\mathcal{Y}^{(2)}$, is encoded by CANF conditioned on $\hat{\mathcal{Y}}^{(1)}$ to produce another bitstream, which is then decoded to produce $\hat{\mathcal{Y}}^{(2)}$. Both $\hat{\mathcal{Y}}^{(1)}$ and $\hat{\mathcal{Y}}^{(2)}$ are then fed to the fusion model to produce $\hat{\mathcal{Z}}$. }
\label{fig:approach2}
\end{figure*}

To compress these features in our proposed system, we utilized two learned image codecs -- ANFIC and CANF: 
\begin{itemize}
    \item \textbf{ANFIC} \cite{anfic} is a high-performance leanrned image codec based on Augmented Normalizing Flows (ANF)~\cite{ae_limit}, in which multiple variational autoencoders (VAE) are stacked to extend the image compression capabilities of the VAE-based image codecs \cite{balle2018}. Normalizing Flows are a class of generative models that use a sequence of invertible transformations to map a simple distribution (e.g., a Gaussian) to a more complex distribution that matches the data distribution \cite{ANF}. Since VAEs are a special case of ANF, it has been argued~\cite{anfic} that ANF-based codecs may theoretically provide more efficient compression.
    \item \textbf{CANF} \cite{canf} is the conditional version of ANFIC, originally designed for conditional inter-frame coding in videos. In fact, while ANFIC adopts ANF to learn the (unconditional) image distribution $p(x)$ for image compression, CANF learns a generative model by maximizing the conditional likelihood $p(x|x_c)$, where $x_c$ is a predictor for $x$. As discussed in \cite{canf,lccm}, conditional coding is theoretically more efficient than the widely-employed residual coding.
\end{itemize}

In our experiments, we considered the following three coding approaches:
\begin{itemize}
\item \textbf{Approach 1}: In this approach, we encoded only the fused features $\mathcal{Z}$ at the output of MAFS. Note that in FUTR3D, the fused feature map has the same size as the individual feature maps at the input of MAFS. For encoding $\mathcal{Z}$, we utilized the ANFIC codec, but we modified it to accept single-channel inputs, with the fused feature map serving as the input. The architecture for this approach is depicted in Fig.~\ref{fig:approach1}.
\item \textbf{Approach 2}: In this approach, we first encoded the camera features $\mathcal{Y}^{(1)}$, followed by the conditional coding of the LiDAR features $\mathcal{Y}^{(2)}$ conditioned on decoded $\widehat{\mathcal{Y}}^{(1)}$. For the camera features, we utilized the ANFIC codec, similar to Approach 1. After coding the camera features, we employed the CANF codec to code the LiDAR features conditioned on the camera features. In our experiments, we first trained the ANFIC codec, followed by the CANF codec. The architecture for this approach is depicted in Fig. \ref{fig:approach2}.
\item \textbf{Approach 3}: This approach is complementary to Approach 2. Here, we first encoded the LiDAR features $\mathcal{Y}^{(2)}$ using ANFIC, and then encoded the camera features $\mathcal{Y}^{(1)}$ conditioned on decoded $\widehat{\mathcal{Y}}^{(2)}$ using CANF. For simplicity, we do not show the figure corresponding to this case. 
\end{itemize}

\begin{figure}
\centering
\includegraphics[width=0.5\textwidth]{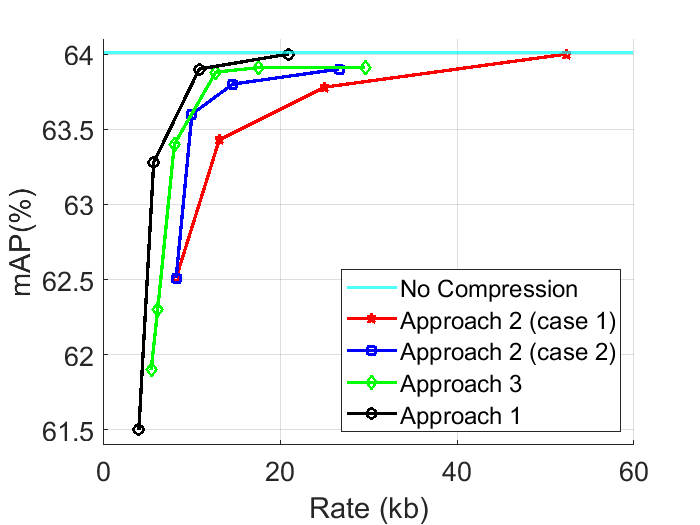}
\caption{The rate-mAP curves for various approaches. }
\label{fig:results}
\end{figure}

\section{Experiments}
\label{sec:experiments}
We evaluated the three coding approaches described in Section~\ref{sec:proposed} on the nuScenes dataset~\cite{nuscenes}. The nuScenes dataset comprises 28,130  samples in the training split and 6,019 samples in the test split. We trained the codecs on the training split and tested them on the test split.

We utilized the original pre-trained FUTR3D network~\cite{futr3d} to produce the features, and then fed the decoded features to the network's back-end. During training, we focused solely on training the codecs while keeping the FUTR3D network parameters frozen. Consequently, the training loss was defined as a Lagrangian rate-distortion loss \cite{anfic}:
\begin{equation}
L = R + \lambda \cdot \text{MSE}(x,\hat{x}),
\end{equation}
where $x$ represents the codec's input (in our case, a feature map), $\hat{x}$ denotes the reconstructed feature map, $R$ is the rate estimate, $\text{MSE}(\cdot,\cdot)$ is the mean squared error, and $\lambda$ is the Lagrange multiplier. In our experiments, we used $\lambda \in \{0.0078125, 0.015625, 0.03125, 0.0625\}$ for all codecs.

For training the ANFIC codecs, we used the procedure defined in~\cite{anfic}, while for training the CANF codec, we followed the procedure outlined in~\cite{canf}. Each experiment comprised a total of 50 training epochs using the Adam optimizer~\cite{adam} with an initial learning rate set to $10^{-4}$.  

We utilized the mean Average Precision (mAP \%)~\cite{futr3d} to measure the object detection accuracy on the test split of nuScenes. The mAP is a popular metric used to evaluate the accuracy of 2D and 3D object detection models. It is computed by first calculating the Average Precision (AP) for each class, which is the area under the Precision-Recall curve. Precision is the ratio of correctly predicted positive samples to all predicted positives, while Recall is the ratio of correctly predicted positives to all actual positives. The mAP is then obtained by averaging the APs across all classes, giving an overall performance score for the model's ability to detect objects across different categories. To measure the rate in Approach 1, we recorded the size (in kilobytes) of the produced compressed bitstream. As a reminder, this bitstream encodes six images and one LiDAR scan. For Approaches 2 and 3, we summed the sizes of the two compressed bitstreams, one for each modality, and considered that as the total rate.

\section{Results and Analysis}
\label{sec:analysis}
The rate-mAP curves for various approaches are shown in Fig.~\ref{fig:results}, where the horizontal cyan line corresponds to the accuracy of FUTR3D without feature compression. Note that for Approaches 2 and 3, two codecs are used, with the second codec (CANF) using the output of the first codec (ANFIC) for conditioning. Since we used four different Lagrange multipliers for each codec, 16 different combinations could be created. To reduce experimental complexity, we examined two specific cases for Approach 2:
\begin{itemize}
    \item \textbf{Case 1}: We used $\lambda=0.0078125$ for ANFIC and $\lambda \in \{0.0078125, 0.015625, 0.03125, 0.0625\}$ for CANF.
    \item \textbf{Case 2}: We used the same $\lambda$ for both codecs. For instance, we trained CANF with $\lambda=0.0625$ using an ANFIC trained with $\lambda=0.0625$.
\end{itemize}
Note that the smaller the value of $\lambda$, the lower the rate. Fig.~\ref{fig:results} shows the results for both cases. 

\begin{figure}[t]
\centering
\includegraphics[width=0.5\textwidth]{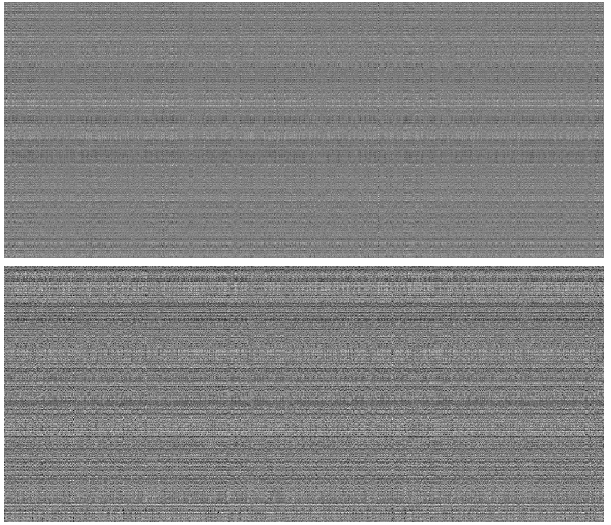}
\caption{A sample feature map for the camera modality (top), and the LiDAR modality (bottom).}
\label{fig:feature_maps}
\end{figure}

As seen in Fig. \ref{fig:results}, Case 2 for Approach 2 yields better results than Case 1. That is, conditioning on the smallest $\lambda$ (lowest rate) gives better results. Consequently, in our experiments, we considered only Case 1 for Approach 3. The results also indicate that Approach 3 -- coding camera features conditioned on LiDAR features -- achieves better rate-accuracy performance than Approach 2 -- coding LiDAR features conditioned on camera features. We attribute this result to the fact that CANF was developed for natural image coding, and camera features tend to smoother and therefore more similar to natural images compared to LiDAR features, as illustrated in Fig.~\ref{fig:feature_maps}. Although we trained CANF within our pipeline to encode the corresponding features, it is possible that the architectural choices made CANF more suitable for image coding rather than coding LiDAR features.

The best rate-accuracy performance is achieved by Approach 1 - coding the fused features. This is not surprising, since the fused feature map $\mathcal{Z}$ is obtained from uni-modal feature maps $\mathcal{Y}^{(1)}$ and $\mathcal{Y}^{(2)}$ via a non-invertible MAFS process. Hence, $I(X;\mathcal{Z}) < I(X;\mathcal{Y}^{(1)},\mathcal{Y}^{(2)})$, and some task-irrelevant information has been removed through MAFS. Therefore, a lower rate is needed for coding $\mathcal{Z}$ compared to coding $\mathcal{Y}^{(1)}$ and $\mathcal{Y}^{(2)}$ in order to achieve the same accuracy. In particular, Approach 3 can achieve the no-compression mAP at a rate of about 21 KBytes. To put this in context, consider that one sample consists of six RGB images of resolution 1600$\times$1900 and a LiDAR scan whose  size is about 678 KBytes. Hence, the raw data size of one sample is around 52 MBytes. Our results show that the no-compression mAP can be achieved with about 21 KBytes per sample, which is almost 2500:1 compression.

We used the Bj\o{}ntegaard Delta Rate (BD-Rate)~\cite{Bjontegaard} to measure the difference between various rate-accuracy curves in Fig.~\ref{fig:results}. We set Approach 2 (case 1) as the anchor, since it yielded the worst results in our experiments, so the BD-Rate results correspond to the average rate savings at the equivalent accuracy compared to this approach. The results are shown in Table~\ref{tab:bd_rate}. As seen here, Approach 1 requires, on average, 67.7\% lower rate than Approach 2 (case 1) to achieve the same 3D object detection accuracy. 

To compare against conventional coding strategies, we performed the following experiment. We encoded images from 1,000 samples of the test split of nuScenes using x265 with QP=51, which gives the lowest rate. The average file size per sample was $\sim$30 KBytes. When decoded images were used with uncompressed LiDAR data, the resulting mAP of FUTR3D was 62.3\%, which is about 1.7\% lower than the no-compression mAP. Meanwhile, our Approach 1 reaches the no-compression mAP with $\sim$21 KBytes, including both camera and LiDAR data. Hence, our Approach 1 requires $\sim$30\% lower rate than images alone coded using x265, while achieving higher accuracy. The key to this superiority, of course, is the removal of task-irrelevant information. We also repeated this experiment using VTM~\cite{vtm} (v23.4), which is the reference software for VVC~\cite{VVC}. The average file size per sample was $19$ KBytes. When decoded images were used with uncompressed LiDAR data, the resulting mAP of FUTR3D was $63.4$\%. With reference to Fig.~\ref{fig:results}, this is still below the performance obtained by all our approaches, despite the fact that uncompressed LiDAR data were used alongside VTM-coded images, while the rate for LiDAR data was not counted.

\begin{table}[t]
    \centering
    \caption{The BD-Rate-mAP results for various approaches relative to Approach 2 (case 1).} 
    \begin{tabular}{c|c|c}
    \toprule
         Approach 1& Approach 3& Approach 2 (case 2)\\
         \midrule
        --67.7\%& --50.4\%& --47.6\%\\
         \bottomrule
    \end{tabular}
    \label{tab:bd_rate}
\end{table}

\begin{table}[b]
    \centering
    \caption{The average on-board execution time (sec) per sample from the nuScenes dataset. } 
    \begin{tabular}{c|c|c|c}
    \toprule
         Use case& Approach 1& Approach 2 & Approach 3\\
         \midrule
       On-board inference & 0.67&  0.95 & 0.92\\
       Edge-cloud inference & 0.28 & 0.48 & 0.48\\
         \bottomrule
    \end{tabular}
    \label{tab:complexity}
\end{table}

Table~\ref{tab:complexity} compares various methods in terms of the average on-board execution time per sample 
on an NVIDIA GeForce RTX 2080 Ti GPU with 12 GB of RAM. We considered two use cases: (1) on-board inference and (2) edge-cloud inference. In on-board inference, object detection is carried out on-board and the purpose of compression is to lower the data rate and make on-board sensor data processing easier. In this case, the entire FUTR3D model and the codec(s) are deployed on board. In edge-cloud inference, only the front-end of FUTR3D is deployed on board, while the back-end is in the cloud. In Approach 1 (Fig.~\ref{fig:approach1}), ANFIC encoder is also employed on board to generate the compressed bitstream to be sent to the cloud. In Approaches 2 and 3 (Fig.~\ref{fig:approach2}), the entire ANFIC codec and a CANF ecoder are deployed on board to generate the two bitstreams that are sent to the cloud. With both approaches, bitstream decoding and object detection are performed in the cloud. As seen in the table, Approach 1 has lower execution time in both use cases, since it requires only one codec. Also, in the edge-cloud use case, all approaches have lower on-board execution time than running the entire FUTR3D model on board without compression, which requires 0.56 seconds/sample.

\section{Conclusions}
\label{sec:conclusions}
In this paper, we explored several learned multimodal compression strategies for autonomous driving. We used a pre-trained network for 3D object detection from camera and LiDAR data to remove task-irrelevant information, and utilized learned codecs to perform multimodal feature compression. We explored several coding approaches, including joint coding of fused modalities and conditional coding of one modality based on the other.  Experimental results on the nuScenes dataset indicate that the joint coding of fused modalities was the most efficient, a result that can be explained in terms of the removal of task-irrelevant information.

{\small
\bibliographystyle{IEEEtran}
\bibliography{ref}
}

\end{document}